\documentclass{ws-ijmpc}
\usepackage[super]{cite} 
\usepackage{epstopdf}
\usepackage{color}

\begin{document}

\markboth{Lijie Cui \& Chuandong Lin}
{Kinetic modeling of economic markets with heterogeneous saving propensities}

\catchline{}{}{}{}{}

\title{Kinetic modeling of economic markets with heterogeneous saving propensities}

\author{Lijie Cui}

\address{International School of Business $\&$ Finance, Sun Yat-sen University,\\ Zhuhai 519082, China}

\author{Chuandong Lin}

\address{Sino-French Institute of Nuclear Engineering and Technology, Sun Yat-sen University,\\ Zhuhai 519082, China \footnote{Corresponding author, e-mail: linchd3@mail.sysu.edu.cn}
}

\maketitle

\begin{history}
\received{Day Month Year}
\revised{Day Month Year}
\end{history}

\begin{abstract}

The lattice gas automaton (LGA) is proposed for a closed economic market of agents with heterogeneous saving interests. There are two procedures in the standard LGA, i.e., ``propagation" + ``transaction". If the propagation step is removed and the transaction is conducted among all agents, the LGA reduces to a more simplified kinetic model. In addition, two dealing rules are imposed on the transaction phase. Under Rule I, the trading volume depends on the average saving propensities of an arbitrary pair of agents in trade. Under Rule II, the exchange is governed by a stochastic parameter between the saving propensities of two traders. Besides, two sampling methods are introduced for the random selection of two agents in the iterative process. Specifically, Sampling I is the sampling with replacement and is easier to program. Sampling II is the sampling without replacement and owns a higher computing efficiency. There are slight differences between the stationary wealth distributions simulated by using the two transaction rules and sampling approaches. In addition, the accuracy, robustness and efficiency of the econophysics models are validated by typical numerical tests. The reduced LGA without the propagation step owns a higher computational efficiency than the standard LGA. Moreover, the impact of saving propensities of agents in two groups on the wealth distributions is studied, and the influence of proportions of agents is investigated as well. To quantitatively measure the wealth inequality, the Gini coefficients, Kolkata indices, and deviation degrees of all agents and two groups are simulated and analyzed in detail. This work is helpful to further analyze and predict the dynamic process of wealth distribution in the realistic economic market.

\end{abstract}

\ccode{PACS Nos.: 89.65.Gh, 89.90.+n}

\section{Introduction}
Econophysics is a typical interdiscipline that applies theories and methods of statistical physics to solve problems in economics \cite{Mantegna1999}. As a fundamental issue in econophysics and economics, wealth distribution plays a key role in human society \cite{Acemoglu2002QJE,Ludwig2022PTRSA}. Due to the great importance, a larger number of economists, mathematicians and physicists are devoted to the scientific research of wealth distribution  \cite{Cordier2005JSP,Aktaev2022PA,Ludwig2022PTRSA}. In theory, a financial society is analogous to a physical system \cite{Cordier2005JSP}. To be specific, the agent, wealth and average wealth per agent are equivalent to the gas molecule, energy and temperature, respectively \cite{Chakraborti2000EPJB}. By analogy with physical models, econophysics modeling of complex social systems has attracted more and more attention \cite{Chakraborti2000EPJB,Quevedo2016CE}. In general, for a coarse-graining financial model, the laws of agent and money conservation are obeyed in a closed market during a certain stage \cite{Quevedo2016CE}. Based on rational assumptions, a series of successful theoretical models and numerical simulations promise deep insights into the characteristics and mechanisms of wealth distributions \cite{Dragulescu2001,Quevedo2016CE,Cardoso2020PA}.

In statistical mechanics, an ideal gas system holds the Boltzmann-Gibbs law, 
${p_i} \propto \exp \left( - {E_i} /\left( k T\right) \right)$, 
where $p_i$ is the probability in state $i$, $E_i$ the corresponding energy, $T$ the thermodynamic temperature, and $k$ the Boltzmann's constant \cite{Gibbs1902}. Similarly, the wealth distribution also takes the Boltzmann-Gibb exponential form in a closed equilibrium system with arbitrary and random trade \cite{Dragulescu2001} (see Eq. \ref{Probability1}) 
\footnote{
	Comparing the mathematical expressions of energy and wealth distributions, it can be found that the agent, wealth ($m$) and average wealth ($m_0$) correspond to the gas molecule, energy ($E_i$) and temperature ($T$), respectively.
}. 
Besides, another famous empirical form for the wealth distribution is the Pareto power-law function, $P\left( m \right)\propto {{m}^{-\alpha }}$, in terms of the Pareto index $\alpha$ and money $m$ \cite{Cardoso2020PA}. Usually, the Boltzmann-Gibbs function fits well the low and middle ranges of wealth distribution \cite{Dragulescu2001,Tao2019JEIC,Cardoso2020PA}, and the Pareto function is in line with the high range \cite{Nirei2007RIW,Newby2011EM,Cardoso2020PA}. Although the classical economic theories are helpful for the study of financial markets, a deeper understanding of the dynamics of trading process often requires more versatile methodologies.

With the rapid development of computational science and technology, numerical simulations serve as a convenient tool for the analysis of economic issues. As an interdisciplinary approach, econophysics modeling of financial markets has made great success in recent decades \cite{Chakraborti2000EPJB,Chatterjee2004PA,Cerda2013MCM,Lima2017PA,Huo2017IJMPC,Pinasco2018DGA,Cui2020Entropy,Cui2021PA,Zhou2021MPE,Liu2021PRE,Paul2022PTRSA}. Early in 2000, the Chakraborti-Chakrabarti (CC) model was proposed for a closed economic system and the influence of saving propensities upon the equilibrium probability distribution of money was studied \cite{Chakraborti2000EPJB}. In 2004, Chatterjee, Chakrabarti and Manna presented the CCM model for an economic market where the interest of saving varies from person to person \cite{Chatterjee2004PA}. In 2017, Lima employed the anisotropic Ising model in an external field and with an ion single anisotropy term as a mathematical model for the financial dynamics \cite{Lima2017PA}. In 2018, Pinasco et al. proposed a model of wealth distribution considering both the simplicity of statistical physics models and the flexibility of evolutionary game theory \cite{Pinasco2018DGA}. In 2021, Zhou et al. introduced the wealth substitution rate into the collision kernel of the Boltzmann equation to investigate the wealth distribution \cite{Zhou2021MPE}. In the same year, Liu et al. simulated economic growth and wealth distribution with a generalized asset exchange model \cite{Liu2021PRE}. Recently, Paul et al. investigated the variations of income or wealth distribution of agents with saving propensities by using kinetic exchange models \cite{Paul2022PTRSA}. 

As an effective kinetic methodology, the lattice gas automaton (LGA) is successfully extended to various fields, including the fluid flow \cite{HPP1973,FHP1986PRL}, chemical reaction \cite{Chen2016Entropy}, metal passivation \cite{Stepien2019EA}, ship evacuation \cite{Han2021MIS}, and wealth distribution \cite{Cerda2013MCM,Cui2020Entropy,Cui2021PA}. In 2013, Cerd$\rm{\acute{a}}$ et al. developed the LGA economic model for income distribution and tax regulation, where economic agents are treated as gas particles moving on a two-dimensional lattice and interacting with each other, and economic transactions are modeled by particle collisions, and money is conserved \cite{Cerda2013MCM}. 
In 2020, Cui and Lin developed an LGA kinetic model of the income distribution under the conditions of the Matthew effect, income tax and charity \cite{Cui2020Entropy}. Next year, Cui and Lin further proposed a simple and efficient one-dimensional LGA for the wealth distribution of agents with or without saving propensities \cite{Cui2021PA}. 
In fact, the LGA can be regarded as the precursor to the powerful mesoscopic methodologies, lattice Boltzmann method \cite{SucciBook,LGABook2005,Lai2014PA,Yan2019CMA} and discrete Boltzmann method \cite{Gan2015SM,Lin2018CNF,Gan2018PRE,Lin2019PRE,Chen2021FOP,Su2022CTP,Ji2022JCP}. Those versatile kinetic models own similar numerical merits (including wide applicability, high accuracy, good robustness, simple scheme, easy programing, and high parallel computing efficiency) and provide convenient tools for scientific and engineering research. 
On the basis of the aforementioned econophysics models \cite{Chatterjee2004PA,Cerda2013MCM,Cui2021PA,Paul2022PTRSA}, here we propose the standard and simplified lattice gas automata, and further investigate the commercial transactions between agents with various saving interests. 

The rest of the paper is organized as follows. In Sec. \ref{SecII}, the economic transaction models and computational approaches are introduced in detail. In Sec. \ref{SecIII}, the numerical robustness and reliability of the kinetic models are validated. Then we study the wealth distributions of two groups of agents with various saving propensities in Sec. \ref{SecIV}. Finally, conclusions are drawn in Sec. \ref{SecV}.
\section{Kinetic method}\label{SecII}

Let us consider a closed economic market, where the total amount of money ${M}$ and the number of agents $N \gg 1$ are constant during a certain period. Consequently, the mean money is ${m_0} = {M} / {N}$. The agent $A_i$ is introduced to represent an individual or corporation, with the subscript $i = 1$, $2$, $\dots$, $N$. The agent ${A_i}$ possesses money ${m}_{i}$ and has a saving propensity $\lambda_{i}$ within the range $0 \le \lambda_{i}\le 1$. The trader saves a fraction of his money $\lambda_{i} {m}_{i}$ and trades randomly with the rest $(1 - \lambda_{i}) {m}_{i}$. The saving interest varies from person to person. 

\subsection{Standard lattice gas automaton}\label{class}
\begin{figure}[htbp]
	\centerline{\includegraphics[width=6cm]{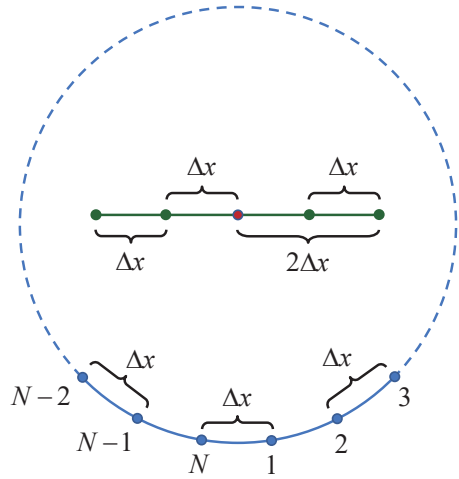}}
	\vspace*{8pt}
	\caption{Computational domain with $N$ sites and one-dimensional four-neighbor model. \label{Fig01}
	}
\end{figure}

To mimic the evolution of the economic society, we present the LGA that consists of two main steps, i.e., ``propagation"  $+$ ``transaction" (or ``collision"). In the propagation step, an agent may transfer randomly to an adjacent empty position (with probability $P_m$) or keep resting (with probability $1 - {P_m}$), which is similar to the physical process where a gas molecule moves or not. Specifically, the location of agent $A_i$ changes from $\mathbf{{X}}$ at time $t$ into $\mathbf{{X}'}$ at time $t + \Delta t$, and the corresponding money is expressed as follows, 
\begin{equation}
	{{m}_{i}}\left( \mathbf{{X}'}, t + \Delta t \right)={{m}_{i}}\left( \mathbf{X},t \right)
	\tt{,}
	\label{Propagation}
\end{equation}
where $\Delta t = 1$ is the temporal step \cite{Cui2020Entropy,Cui2021PA}. 
The distance between $\mathbf{{X}'}$ and $\mathbf{X}$ is relevant to the spatial step $\Delta x = 1$ \cite{Cui2020Entropy,Cui2021PA}. To be specific, for the two-dimensional four-neighbor model, $|\mathbf{{X}'} - \mathbf{X} | = 0$ or $\Delta x$, namely, an agent either remains motionless or propagates to one of its four adjacent vacant sites randomly \cite{Cui2020Entropy}. In the two-dimensional eight-neighbor model, $|\mathbf{{X}'} - \mathbf{X} | = 0$, $\Delta x$, or $\sqrt{2} \Delta x$, that is to say, each agent has the opportunity to move to one of its nearest four sites or the four diagonal sites \cite{Cui2020Entropy}. 
Similarly, for the one-dimensional four-neighbor model, $|\mathbf{{X}'} - \mathbf{X} | = 0$, $\Delta x$, or $2\Delta x$, i.e., the agent may stay or leave to an empty grid point \cite{Cui2021PA}. Figure \ref{Fig01} delineates the computational domain with $N$ sites and one-dimensional four-neighbor model \cite{Cui2021PA}. Without loss of generality, only the one-dimensional four-neighbor model is employed for the numerical simulations in this paper. 

In the transaction stage, two neighboring agents exchange money with a certain probability $P_t$ (in analogy to the collision process of an ideal gas system) \cite{Cerda2013MCM,Cui2020Entropy,Cui2021PA}. For an arbitrary pair of agents $A_i$ and $A_j$ in trade, their money changes as follows
\begin{equation}
	\left\{
	\begin{array}{l}
		{{m}'_{i}} = {{m}_{i}} - \Delta m \tt{,}  \\
		{{m}'_{j}} = {{m}_{j}} + \Delta m \tt{.}
	\end{array}
	\right.
	\label{ExchangeM}
\end{equation}
where $\Delta m$ stands for the trading volume, ${{m}_{i}}$ and ${{m}'_{i}}$ denote the money of $A_i$ before and after the transaction, respectively. ${{m}_{j}}$ and ${{m}'_{j}}$ represent the money of $A_j$ before and after the transaction, respectively. For simplicity, no debt or profit is under consideration in this work. 

Actually, the key is to determine the mathematical expression of $\Delta m$ in Eq. \ref{ExchangeM}. As for the CCM model in Ref.~\refcite{Chatterjee2004PA}, each money-conserving trading is identified as an elastic two-body collision, and the trading volume is expressed by
\begin{equation}
	\Delta m=\left(1-{{\lambda }_{i}} \right) \left(1-\varepsilon \right){{m}_{i}} - \varepsilon \left(1-{{\lambda }_{j}} \right){{m}_{j}}
	\label{DeltaM_CCM}
	\tt{,}
\end{equation}
where $\varepsilon$ is a random fraction uniformly distributed in the interval $0 \le \varepsilon \le 1$, coming from the stochastic nature of the trading \cite{Chakraborti2000EPJB,Chatterjee2004PA}. Substituting Eq. \ref{DeltaM_CCM} into Eq. \ref{ExchangeM} results in
\begin{equation}
	\left\{
	\begin{array}{l}
		{{m}'_{i}} = {{\lambda}_{i}} {{m}_{i}} + \Delta {m_i} \tt{,}  \\
		{{m}'_{j}} = {{\lambda}_{j}} {{m}_{j}} + \Delta {m_j} \tt{,}
	\end{array}
	\right.
	\label{ExchangeM_Form2}
\end{equation}
with 
\begin{equation}
	\left\{
	\begin{array}{l}
	{\Delta {m_i}} = \varepsilon \left(1-{{\lambda }_{i}} \right){{m}_{i}}+\varepsilon \left(1-{{\lambda }_{j}} \right){{m}_{j}} \tt{,}   \\
	{\Delta {m_j}} = \left(1-\varepsilon \right) \left(1-{{\lambda }_{i}} \right){{m}_{i}}+ \left(1-\varepsilon \right) \left(1-{{\lambda }_{j}} \right){{m}_{j}}  \tt{.}
    \end{array}
    \right.
    \label{DeltaM_CCM_Form2}
\end{equation}
The meaning of Eqs. \ref{ExchangeM_Form2} and \ref{DeltaM_CCM_Form2} is as follows. Trader $A_i$ ($A_j$) saves a random fraction $\lambda_i$ ($\lambda_j$), and his money changes by amounts ${\Delta {m_i}}$ (${\Delta {m_j}}$) during the transaction. Here ${\Delta {m_i}}$ is a random fraction $\varepsilon$ of $(1-{{\lambda }_{i}}){{m}_{i}}+ (1-{{\lambda }_{j}}){{m}_{j}}$ and ${\Delta {m_j}}$ is the rest ($1-\varepsilon$) of it. That is to say, the conservation of money is ensured in each trade \cite{Chatterjee2004PA}. 
Although the trading rule is based on the physical foundation of an elastic collision process, this exchange model predicts that an agent with a smaller saving propensity will have less and less wealth until no money is left \cite{Chatterjee2004PA}, which is far from the actual economic situation, see \ref{AppendixA}. This fatal defect hinders the reliable simulation and analysis of an economic market. 

For the sake of a realistic prediction of monetary transactions, we propose the trading volume as follows, 
\begin{equation}
	\Delta m=\left( 1-{{\lambda }_{e}} \right)\left[ {{m}_{i}}-\varepsilon \left( {{m}_{i}}+{{m}_{j}} \right) \right]
	\label{DeltaM}
	\tt{,}
\end{equation}
where the parameter ${\lambda }_{e}$ depends on the saving propensities $\lambda_{i}$ and $\lambda_{j}$. In this paper, we present two types of transaction rules, i.e., Rule I and Rule II. To be specific, the parameter is ${\lambda }_{e} = ({\lambda}_{i}+{\lambda}_{j})/2$ for Rule I, and ${\lambda }_{e}$ is a stochastic number within the range, $\min \left(\lambda_{i}, \lambda_{j} \right) \le {\lambda }_{e} \le \max \left(\lambda_{i}, \lambda_{j} \right)$ for Rule II. 
It should be mentioned that, with Eq. \ref{DeltaM}, the formula in Eq. \ref{ExchangeM} could be written as
\begin{equation}
		\left\{
		\begin{array}{l}
			{{m}'_{i}} = {{\lambda}_{e}} {{m}_{i}} + \Delta {m_i} \tt{,}  \\
			{{m}'_{j}} = {{\lambda}_{e}} {{m}_{j}} + \Delta {m_j} \tt{,}
		\end{array}
		\right.
\end{equation}
in terms of 
\begin{equation}
		\left\{
		\begin{array}{l}
			\Delta {{m}_{i}}=\varepsilon \left( 1-{{\lambda }_{e}} \right)\left( {{m}_{i}}+{{m}_{j}} \right)  \tt{,}   \\
			\Delta {{m}_{j}}=\left( 1-\varepsilon  \right)\left( 1-{{\lambda }_{e}} \right)\left( {{m}_{i}}+{{m}_{j}} \right)  \tt{.}
		\end{array}
		\right.
\end{equation}

In fact, the above two mathematical expressions are sensible. In a real-life scenario, an agreement is made between two businessmen with different saving propensities $\lambda_{i}$ and $\lambda_{j}$. After discussion between them, the two parties achieve an agreement that the same wealth fraction $\lambda_{e}$ is in trade, where $\lambda_{e}$ depends upon $\lambda_{i}$ and $\lambda_{j}$. In addition, it is clear that the agent $A_i$ is a ``loser" and the other agent $A_j$ becomes the ``winner" in the case of $\Delta m > 0$, and vice versa. Moreover, it can be found from Eq. \ref{DeltaM_CCM} or \ref{DeltaM} that the relation ${{m}_{i}} \ge {\Delta m}$ 
(or ${{m}_{j}} \ge -{\Delta m}$) 
is satisfied when the agent $A_i$ (or $A_j$) is a ``loser". Consequently, the debt does not occur because the relations ${{m}'_{i}} \ge 0$ and ${{m}'_{j}} \ge 0$ are satisfied in Eq. \ref{ExchangeM}.

\subsection{Reduced lattice gas automaton}

It is interesting to note that the standard LGA reduces to a more simplified kinetic model if there is no propagation step and all agents can exchange with each other randomly. Here the simplified kinetic model is also named the reduced LGA. 

In fact, how to choose two agents with random transactions is a key to the reduced LGA. In this paper, two sampling methods (i.e., Samplings I and II) are introduced for the random selection of two agents in trade. To be specific, Sampling I is sampling with replacement. Each agent has an equal chance to be selected, and a pair of agents are chosen from all samples randomly with the same probability at each iterative step in the loop of computer programming. 

\begin{figure}[htbp]
	\centerline{\includegraphics[width=10cm]{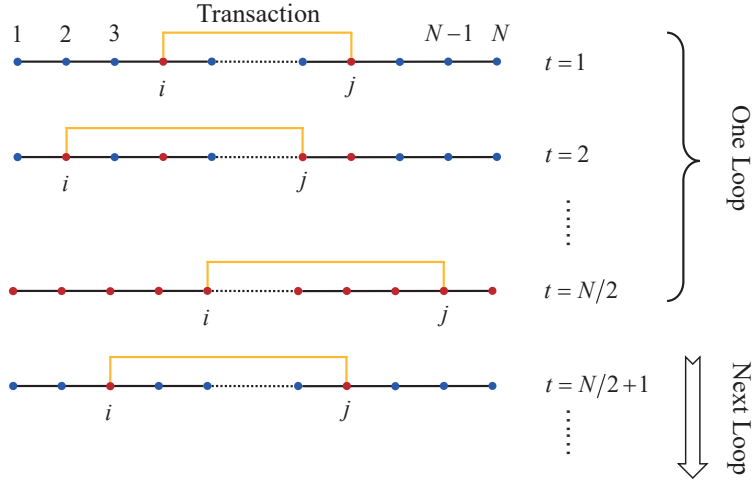}}
	\vspace*{8pt}
	\caption{
			Sketch of sampling without replacement. \label{Fig02}
	}
\end{figure}

On the other hand, Sampling II is sampling without replacement. Namely, we deliberately avoid choosing an agent of the sample more than once during one main loop that takes $N/2$ temporal steps. Specifically, one pair of agents ${A_i}$ and ${A_j}$ are randomly selected from $N$ agents at the first temporal step $t = 1$. Then at the second temporal step $t = 2$, another two agents ${A_i}$ and ${A_j}$ are randomly selected from the remaining $N-3$ samples. In succession, all agents have made a deal once at the end of the main loop at $t = N/2$. Then the next loop begins when $t = N/2+1$, and ends until all agents have made two transactions at $t = N$. Figure \ref{Fig02} displays the sketch of the random selection process of Sampling II. 

It should be mentioned that both Samplings I and II can be adopted to predict the stationary state of the economic market. Actually, the simulated relaxation processes in the evolution of the market are different by using the two sampling methods. In other words, Sampling I takes longer relaxation time than Sampling II, see Fig. \ref{Fig03} for more details. 

In addition, both Rules I and II reduce to the CC model if all agents have an identical saving propensity \cite{Chakraborti2000EPJB}. Besides, in Ref.~\refcite{Chakraborti2000EPJB}, an arbitrary pair of agents $A_i$ and $A_j$ are chosen randomly from all agents $N$ to get engaged in a trade, namely, Sampling I is adopted for the CC model. Consequently, in the case of $\lambda_{i} = \lambda_{j}$, Rule I or II plus Sampling I is equivalent to the CC model, as shown in Fig. \ref{Fig03}. 

\section{Verification and Validation}\label{SecIII}

To verify and validate the kinetic method for economic markets, three subsections are included here. The first part is for simulations of the relaxation process of the economic market by using the standard LGA and reduced LGA. Next, comparisons are made between exact solutions and numerical results for the transaction without saving propensities. In the third part, the reliability and robustness of kinetic methods are validated via simulations of transactions with saving propensities. 

\subsection{Relaxation process of the economic market}

The first four problems to be resolved are as follows: (i) As the economic markets are continuously traded, whether a steady state could be reached after a sufficiently long period? (ii) What is the difference between the standard and reduced lattice gas automata? (iii) Whether the kinetic models and sampling methods produce a similar steady state of the economic market? (iv) How do the rules and samplings affect the relaxation process of the economic market? For this purpose, numerical simulations are carried out by using the standard LGA and reduced LGA with different dealing rules and sampling methods. 

To measure the wealth inequality in the dynamic process of an economic market, the Gini coefficient for individual agents is defined as 
\begin{equation}
	{\rm{G}} = \frac{1}{2N^{2}{{w}_{0}}}\sum\nolimits_{i=1}^{{{N}_{w}}}{\sum\nolimits_{j=1}^{{{N}_{w}}}{\left| {{w}_{i}}-{{w}_{j}} \right|}}
	\tt{,}
\end{equation}
which depends on the parameters $w_{0} = m_{0}$, $N_{w} = N$, and $w_{i} = m_{i}$. In addition, the Gini coefficient $\rm{G}'$ can also be introduced for two-earner families in terms of $w_{0} = 2 m_{0}$, $N_{w} = N / 2$, and $w_{i} = m_{i}+m_{i+N/2}$. 
The Gini coefficient is an index for the degree of inequality in a social system, used to estimate how far the wealth/income distribution deviates from an equal distribution. 
In theory, both Gini coefficients $\rm{G}$ and $\rm{G}'$ ranges from zero (complete equality) to one (complete inequality). 
With the increasing degree of inequality in monetary distribution, the Gini coefficient increases. 

\begin{figure}[htbp]
	\centerline{\includegraphics[width=12.7cm]{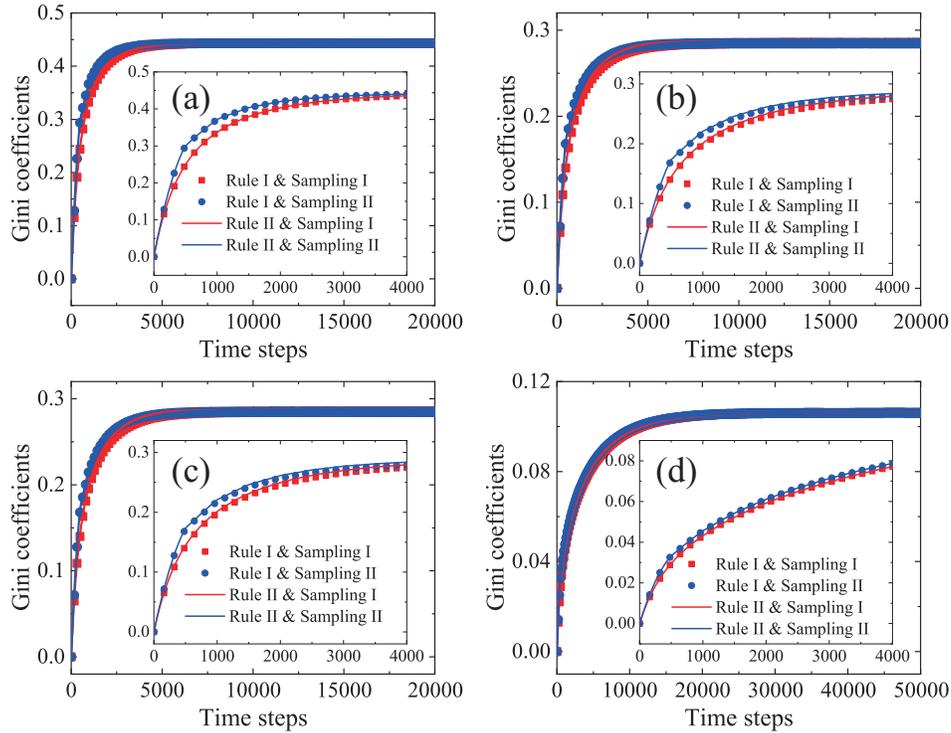}}
	\vspace*{8pt}
	\caption{Evolution of Gini coefficients in four cases: (a) ${\lambda}_{A} = {\lambda}_{B} = 0.1$, (b) ${\lambda}_{A} = 0.3 \ \& \ {\lambda}_{B} = 0.7$, (c) ${\lambda}_{A} = 0.7 \ \& \ {\lambda}_{B} = 0.3$, and (d) ${\lambda}_{A} = {\lambda}_{B} = 0.9$. The insets show the corresponding profiles within $0 \le t \le 4000$. The squares stand for the simulation results of Rule I $\&$ Sampling I, the circles for Rule I $\&$ Sampling II, the red lines for Rule II $\&$ Sampling I, and the blue lines for Rule II $\&$ Sampling II. \label{Fig03}}
\end{figure}

Now, let us consider the following configuration. There are $N = 1000$ agents in a closed economic market, and each agent initially possesses the same money $m_i = 1$. Half agents have the same saving propensity $\lambda_{A}$, and the other traders own the saving propensity $\lambda_{B}$. Specifically, four cases of the saving propensities are considered: (a) ${\lambda }_{A} = {\lambda }_{B} = 0.1$, (b) ${\lambda}_{A} = 0.3 \ \& \ {\lambda}_{B} = 0.7$, (c) ${\lambda}_{A} = 0.7 \ \& \ {\lambda}_{B} = 0.3$, and (d) ${\lambda}_{A} = {\lambda}_{B} = 0.9$. Figure \ref{Fig03} illustrates the evolution of Gini coefficients for individual agents in the four cases. From Fig. \ref{Fig03}, the following points can be obtained.

(i) The Gini coefficient increases from zero to a steady value, and its change rate reduces gradually in the evolution. Namely, the economic system reaches the steady state as there are continuous trades over a sufficiently long period. 
The initial Gini index $\rm{G} = 0$ expresses perfect equality when each agent possesses the same wealth. After a series of transactions, the wealth distribution deviates from the initial equal distribution, and the degree of inequality increases gradually until the final equilibrium state is achieved. 
	\footnote{
	It is interesting to point out that the relaxation process of economic transaction is analogous to the thermodynamic relaxation process. Suppose that there is a macroscopic physical system where each gas molecule owns the same energy initially. The velocity (or energy) distribution changes step by step due to a sequence of collisions, and the equilibrium state emerges after the relaxation period in the system.
	}

(ii) With the increasing of saving propensities ${\lambda}_{A}$ and ${\lambda}_{B}$, the steady values of the Gini coefficients decrease. To be specific, the steady Gini coefficients are about ${\rm{G}_{S}} = 0.443$, $0.285$, $0.285$, and $0.106$ in Figs. \ref{Fig03} (a)-(d), respectively. In other words, the inequality of income or wealth reduces as the saving propensities increase, which is reasonable \cite{Chakraborti2000EPJB,Chatterjee2004PA}. 

(iii) There is a remarkable agreement among the final steady Gini indexes obtained from the simulations by using Rule I $\&$ Sampling I, Rule I $\&$ Sampling II, Rule II $\&$ Sampling I, and Rule II $\&$ Sampling II. That is to say, Rules (I and II) and Samplings (I and II) give a similar steady state of the economic market. 

(iv) As shown in the insets, the simulated Gini coefficients by using Rule I $\&$ Sampling I and Rule II $\&$ Sampling I coincide with each other. Similarly, there is a nice agreement between the numerical results from Rule I $\&$ Sampling II and Rule II $\&$ Sampling II. Namely, there are few differences between Rules I and II in the relaxation process. 

(v) The insets further show that the Gini coefficients given by Rule I $\&$ Sampling I are lower than those by Rule I $\&$ Sampling II. Similarly, the Gini coefficients obtained from Rule II $\&$ Sampling I are smaller than those from Rule II $\&$ Sampling II. In fact, the economic market evolves faster in the way of Sampling II than I. From a statistical point of view, to ensure that all agents have business dealings, it takes a longer time by using Sampling I than II due to the different random selection processes. 
Specifically, as shown in Fig. \ref{Fig02}, all agents have made a deal once at the end of the main loop at $t = N/2$, and have made two transactions at $t = N$ for Sampling II. By contrast, for Sampling I, in all likelihood, some agents may have not conducted a transaction after the first or even second main loop. In practice, the economic market achieves the stationary state as long as almost all agents have made multiple transactions. 

Next, to have a quantitative study of the relaxation time, let us introduce a symbol $\tau$ that denotes the time instant when the Gini coefficient increases to $0.99$ times the steady Gini coefficients, i.e., ${\rm{G}} (\tau) = 0.99 {\rm{G}_{S}}$. It can be found that, by using Rule I $\&$ Sampling I, the corresponding temporal steps are $\tau = 4670$, $5930$, $5930$, and $21160$ in Figs. \ref{Fig03} (a)-(d), respectively. The relaxation time becomes longer for larger saving propensities, because it takes a longer time for the market to reach a steady state as the agents transact with less money each time. The simulation results are reasonable \cite{Chakraborti2000EPJB,Chatterjee2004PA}. 

\begin{table}[ht]
	\tbl{Comparison of computing time for various kinetic methods.}
	{\begin{tabular}{@{}cccc@{}} \toprule
			Kinetic method & Time steps ($\Delta t$) & Computing time (s)  \\  \colrule
			\hphantom{00}Standard LGA \hphantom{00} & \hphantom{0}357000 & \hphantom{0}2408 \\ 
			Rule I $\&$ Sampling I \hphantom{00} & \hphantom{0}4670 & \hphantom{0}7.42 \\
			\hphantom{0}Rule I $\&$ Sampling II \hphantom{00} & \hphantom{0}3680 & \hphantom{0}3.89\\
			\hphantom{0}Rule II $\&$ Sampling I \hphantom{00} & \hphantom{0}4670 & \hphantom{0}7.61\\
			\hphantom{00}Rule II $\&$ Sampling II \hphantom{00} & \hphantom{0}3680 & \hphantom{0}5.69 \\ 
			\botrule
		\end{tabular}
		\label{ta1}}
\end{table}

To compare the computational efficiency of the abovementioned methods, table \ref{ta1} displays the time steps and the computing time when ${\rm{G}} (\tau) = 0.99 {\rm{G}_{S}}$. The computational facility is a personal computer with Intel(R) Core(TM) i9-10885H CPU @ 2.40GHz and RAM 64.00 GB. Numerical tests show that the time step is $\tau = 357000$ for the standard LGA, and the time steps are $4670$, $3680$, $4670$, and $3680$ for the reduced LGA with Rule I $\&$ Sampling I, Rule I $\&$ Sampling II, Rule II $\&$ Sampling I, and Rule II $\&$ Sampling II, respectively. And the corresponding computing time is $2408$ seconds for the standard LGA, and $7.42$, $3.89$, $7.61$, and $5.69$ seconds for the simplified model in the four ways, respectively. It is clear that the standard LGA takes much longer time than the reduced LGA. To be specific, the computing time required by the standard LGA is $619$ times the time taken by the simplified model with Rule I $\&$ Sampling II. Consequently, the kinetic model without the propagation phase owns a much higher computational efficiency than the standard LGA. In addition, Rule I takes a little more computing time than Rule II. Sampling II has less computational time than Sampling I. 

\begin{figure}[htbp]
	\centerline{\includegraphics[width=12.7cm]{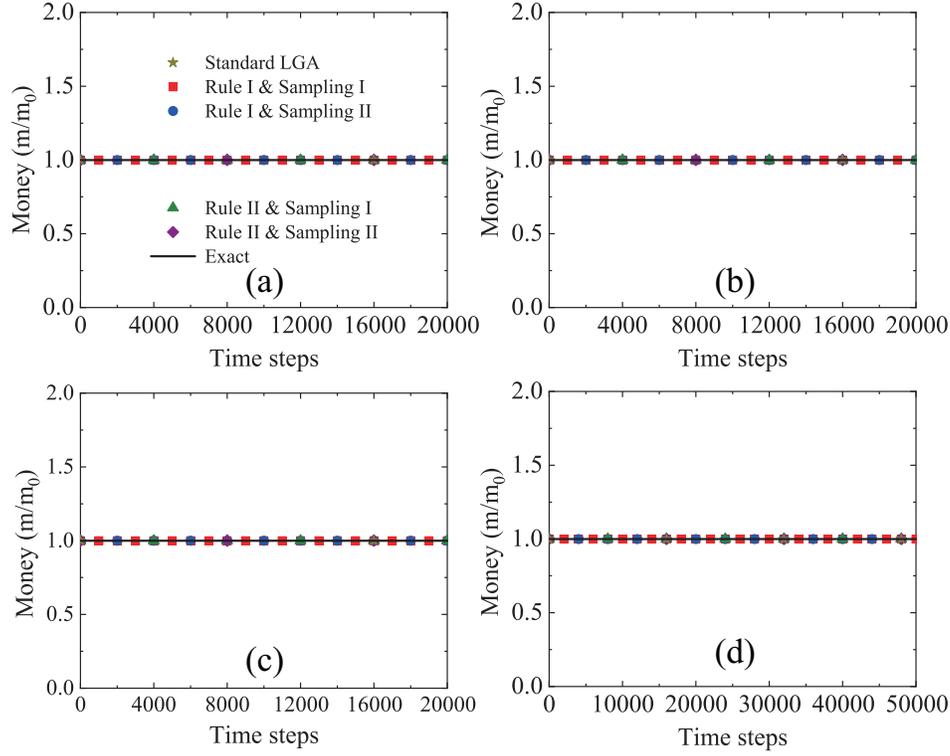}}
	\vspace*{8pt}
	\caption{Evolution of average money in four cases: (a) ${\lambda}_{A} = {\lambda}_{B} = 0.1$, (b) ${\lambda}_{A} = 0.3 \ \& \ {\lambda}_{B} = 0.7$, (c) ${\lambda}_{A} = 0.7 \ \& \ {\lambda}_{B} = 0.3$, and (d) ${\lambda}_{A} = {\lambda}_{B} = 0.9$. The stars stand for the simulation results of the standard LGA, the squares for Rule I $\&$ Sampling I, the circles for Rule I $\&$ Sampling II, the triangles for Rule II $\&$ Sampling I, and the diamonds for Rule II $\&$ Sampling II. The lines denote the exact values.\label{Fig04}}
\end{figure}

In theory, the total money of all agents should be unchanged in the evolution of a closed economic market. To further validate that the standard and reduced lattice gas automata follow the law of money conservation, let us measure the amount of money in the process of economic transactions. Figure \ref{Fig04} displays the average money versus the time in the same four cases as Fig. \ref{Fig03}. Obviously, the mean money remains unchanged during the evolution of economic transactions, as shown in Figs. \ref{Fig04} (a)-(d). That is to say, the principle of money conservation is obeyed by both the standard LGA and the reduced LGA with various transaction rules and sampling techniques. 

\subsection{Transaction without saving propensities}
In theory, it is demonstrated by Dr$\rm{\breve{a}}$gulescu and Yakovenko in Ref.~\refcite{Dragulescu2001} that the monetary distributions of individual agents and two-earner families in the equilibrium state of an ideal free market read
\begin{equation}
	{{P}_{1}}\left( m \right)=\frac{1}{{{m}_{0}}}\exp \left( -\frac{m}{{{m}_{0}}} \right)
	\label{Probability1}
	\tt{,}
\end{equation}
and 
\begin{equation}
	P_{2} \left( m \right) = \frac{m}{m_{0}^{2}} \exp \left(- \frac{m}{m_{0}} \right)
	\label{Probability2}
	\tt{,}
\end{equation}
respectively. 
Obviously, Eq. \ref{Probability1} leads to the relationship ${{P}_{1}}\left( {{m}_{X}} \right){{P}_{1}}\left( {{m}_{Y}} \right)={{P}_{1}}\left( {{m}_{X}}+{{m}_{Y}} \right)$,
which corresponds to the Markovian nature of the scattering or trading processes \cite{Chatterjee2004PA}. 

In addition, the cumulative fraction of individual agents $x_1$ and the cumulative share of money $y_1$ satisfy the following relation \cite{Dragulescu2001},
\begin{equation}
	\left\{
	\begin{array}{l}
		{x_1} = 1-\exp \left( -\frac{m}{{{m}_{0}}} \right) \tt{,}  \\
		{y_1} = {x_1} - \frac{m}{{{m}_{0}}}\exp \left( -\frac{m}{{{m}_{0}}} \right) \tt{.}
	\end{array}
	\right.
	\label{Lorenz1}
\end{equation}
The cumulative fraction of two-earner families $x_2$ and the cumulative share of wealth $y_2$ are expressed by,
\begin{equation}
	\left\{
	\begin{array}{l}
		{x_2} = 1-\left( 1+\frac{m}{{{m}_{0}}} \right)\exp \left( -\frac{m}{{{m}_{0}}} \right) \tt{,}  \\
		{y_2} = {x_2} - \frac{1}{2} \frac{m^{2}}{{{m}_{0}^{2}}} \exp \left( -\frac{m}{{{m}_{0}}} \right) \tt{.}
	\end{array}
	\right.
	\label{Lorenz2}
\end{equation}
\begin{figure}[htbp]
	\centerline{\includegraphics[width=12.7cm]{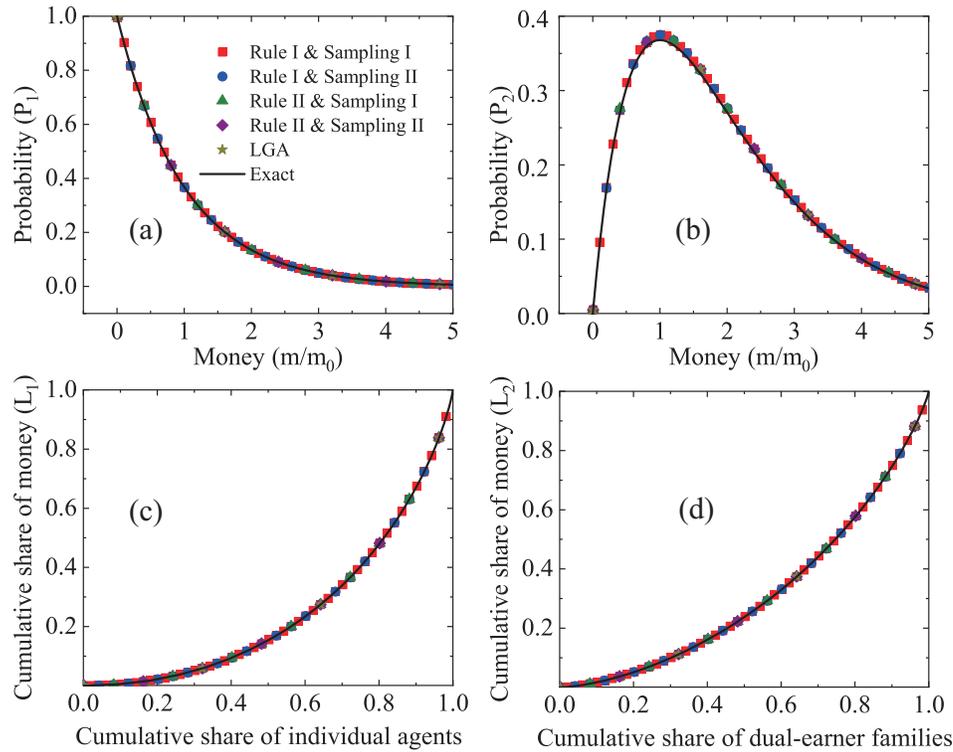}}
	\vspace*{8pt}
	\caption{Money distributions of individual agents  (a) and dual-earner families (b), and cumulative money shares of individual agents (c) and dual-earner families (d) in the case without saving propensities. The stars represent the simulation results of the standard LGA, and the squares, circles, triangles, diamonds are for Rule I $\&$ Sampling I, Rule I $\&$ Sampling II, Rule II $\&$ Sampling I, and Rule II $\&$ Sampling II, respectively. The lines indicate the corresponding exact solutions. \label{Fig05}}
\end{figure}

Next, to verify the kinetic models and sampling techniques, let us make a comparison between the abovementioned theoretical solutions and the numerical results. For the simulations, the saving propensities are ${{\lambda}_{A}} = {{\lambda}_{B}} = 0$, and the other parameters are the same as those in Fig. \ref{Fig04}. Figures \ref{Fig05} (a) and (b) display the money distributions of individual agents and dual-earner families, respectively. Figures \ref{Fig05} (c) and (d) illustrate the cumulative money shares of individual agents and dual-earner families, respectively. The simulation results are indicated by the symbols, and the corresponding exact solutions in Eqs. \ref{Probability1} - \ref{Lorenz2} are represented by the solid lines in Figs. \ref{Fig05} (a)-(d), respectively. 

It is clear in Fig. \ref{Fig05} that all simulation results of the standard LGA and simplified model are in nice agreement with the theoretical results. To be specific, as shown in Fig. \ref{Fig05} (a), the stationary money distribution function becomes the Boltzmann-Gibbs distribution when the agents do not save. It can be found in Fig. \ref{Fig05} (b) that the peak of the curve is located at $m/{m_0} = 1$, namely, the probability of two-earner families achieves the maximum when the wealth of the family equals the average wealth of individual agents. Moreover, the standard LGA gives the simulated Gini coefficients $\rm{G} = 0.495$ and $\rm{G}' = 0.358$ for the individual agents and dual-earner families, respectively. Compared with the exact solutions $\rm{G} = 1/2$ and $\rm{G}' = 3/8$  in Ref.~\refcite{Dragulescu2001}, the relative errors are $1 \%$ and $4.5 \%$, respectively. In addition, the simplified model with Rule I, Rule II, Sampling I, or Sampling II gives similar results. The simulation results are satisfactory. 

\subsection{Transaction with saving propensities}

\begin{figure}[htbp]
	\centerline{\includegraphics[width=12.7cm]{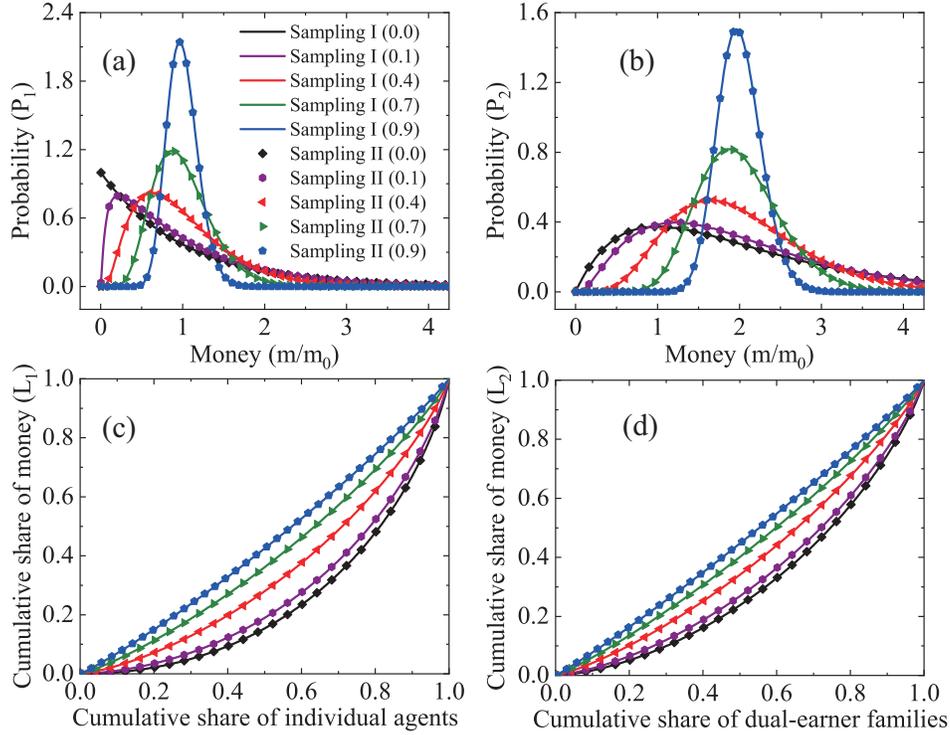}}
	\vspace*{8pt}
	\caption{Money distributions of individual agents (a) and dual-earner families (b), and cumulative money shares of individual agents (c) and dual-earner families (d). The symbols represent the results of Sampling II with ${{\lambda}_{A}} = {{\lambda}_{B}} = 0$ (squares), $0.1$ (sexangles), $0.4$ (left triangles), $0.7$ (right triangles) and $0.9$ (pentagons), respectively. The lines represent the corresponding results of Sampling I.\label{Fig06}}
\end{figure}

\begin{figure}[htbp]
	\centerline{\includegraphics[width=12.7cm]{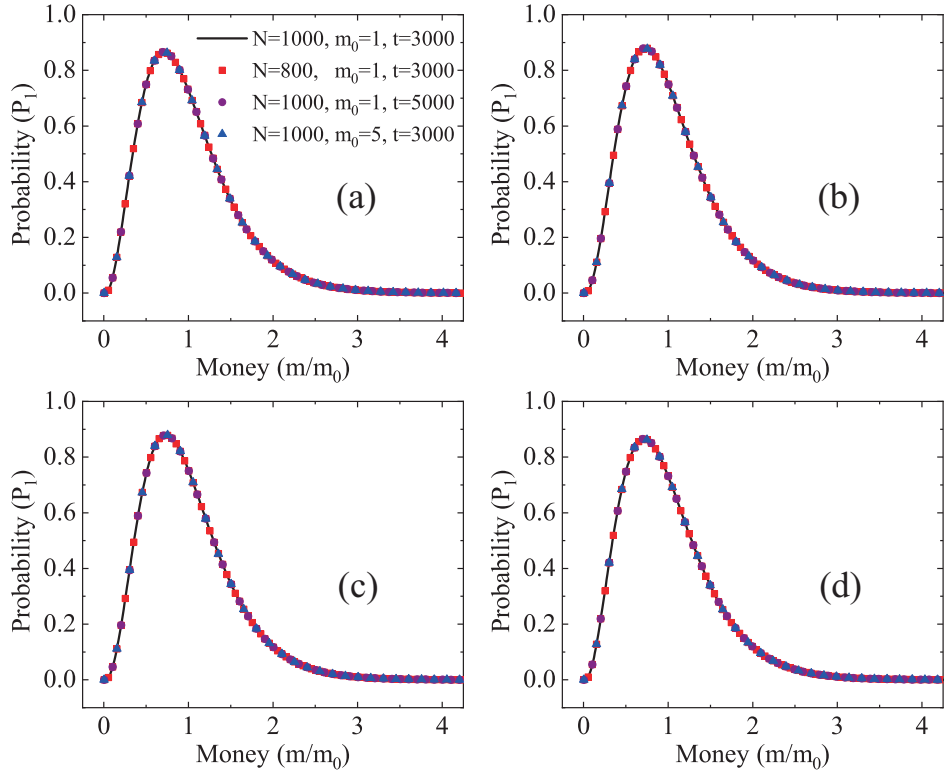}}
	\vspace*{8pt}
	\caption{Money distributions of individual agents given by Rule I $\&$ Sampling I (a), Rule I $\&$ Sampling II (b), Rule II $\&$ Sampling I (c), and Rule II $\&$ Sampling II (d), respectively. The lines and symbols represent simulation results by using different parameters in the legend.\label{Fig07}}
\end{figure}

The main advantage of the dealing rules is the capability of describing monetary transactions with homogeneous or inhomogeneous saving propensities. In the above subsection, it is demonstrated that both the standard LGA and reduced LGA with either of the two transaction rules and sampling methods are suitable for an ideal free market without saving propensities. Next, let us validate that the transaction with saving propensities can also be described in Figs. \ref{Fig06} and \ref{Fig07}. For the sake of brevity, we focus on the simplified kinetic model with the two transaction rules and sampling approaches. The standard LGA has the same performance characteristics and presents identical simulation results (not shown here). 

Figure \ref{Fig06} illustrates the money distributions of individual agents (a) and dual-earner families (b), and displays the cumulative shares of individual agents (c) and dual-earner families (d) in five cases of saving propensities, i.e., ${{\lambda}_{i}} = 0.0$, $0.1$, $0.4$, $0.7$, and $0.9$, respectively. For each case, two agents who get engaged in trade have the same saving propensities, hence both Rules I and II reduce to the CC model \cite{Chakraborti2000EPJB}. For brevity, only Rule I is employed in Fig. \ref{Fig06}, as Rule II and CC model give exactly the same simulation results (not shown here). Actually, the simulated stationary money distributions of Samplings I and II also overlap each other, as shown in Fig. \ref{Fig06}. 

It is evident in Fig. \ref{Fig06} (a) that the stationary money distribution is a function of saving propensities. When all agents do not save, the stationary wealth distribution takes the Boltzmann-Gibb exponential form, with the characteristic of non-interacting agents \cite{Chakraborti2000EPJB,Chatterjee2004PA}. 
Introduction of a finite amount of saving propensities (${\lambda_i} > 0$), dictated by individual self-interest, immediately makes the money dynamics cooperative and the resulting asymmetric Gaussian-like stationary distribution acquires global ordering properties \cite{Chakraborti2000EPJB,Chatterjee2004PA}. 
The money distributions of individual agents ${P_{1}}$ change from the Boltzmann-Gibb form to the asymmetric Gaussian-like form with the increasing saving propensities. This type of self-organization in the economic market comes from pure self-interest of each agent \cite{Chakraborti2000EPJB}.

Besides, Fig. \ref{Fig06} (b) shows that there is a peak of the money distribution of dual-earner families ${P_{2}}$ for any value of ${\lambda}_{i}$, and the peak becomes higher for a larger saving propensity. Moreover, with the increasing ${\lambda}_{i}$, the probabilities of the individual agents or dual-earner families whose money is close to the respective average value gradually grow, and the peaks move rightwards and become thinner and higher. 

Moreover, it can be seen in Figs. \ref{Fig06} (c) and (d) that the ranges of cumulative shares of money are $0 \le {L_{1}} \le 1$ for individual agents and $0 \le {L_{2}} \le 1$ for dual-earner families. Namely, for both individual agents and dual-earner families, the cumulative shares of money rise from zero to one. With the increasing saving propensity, the curve of the cumulative share of money becomes close to the line of perfect equality, and the corresponding curvature decreases. Obviously, Figs. \ref{Fig06} (c) and (d) are consistent with Figs. \ref{Fig06} (a) and (b), and all simulation results are reasonable \cite{Chakraborti2000EPJB,Chatterjee2004PA}. Additionally, it can be found in Figs. \ref{Fig06} (a)-(d) that the wealth distributions are functions of the saving propensities, i.e., 
$P_1 = P_1 (\lambda_A, \lambda_B)$, 
$P_2 = P_2 (\lambda_A, \lambda_B)$, 
$L_1 = L_1 (\lambda_A, \lambda_B)$, 
and 
$L_2 = L_2 (\lambda_A, \lambda_B)$, 
which can be used for the definition of deviation degree in Eq. \ref{Deltaf}. 

To further test the numerical robustness and reliability of the rules and samplings, let us carry out simulations with different numbers of agents $N$, values of average money $m_0$, and time instants $t$. 
The saving propensities are chosen as $\lambda_{A} = \lambda_{B} = 0.4$. 
Figures \ref{Fig07} (a)-(d) display the stationary money distributions of individual agents simulated by using Rule I $\&$ Sampling I, Rule I $\&$ Sampling II, Rule II $\&$ Sampling I, and Rule II $\&$ Sampling II, respectively. The lines, squares, circles, and triangles stand for the simulation results with 
($N$, $m_0$, $t$) $=$ 
($1000$, $1$, $3000$), 
($800$, $1$, $3000$), 
($1000$, $1$, $5000$), and 
($1000$, $5$, $3000$), respectively. Clearly, all numerical results with various parameters are in nice agreement with each other in each plot. Besides, the simulation results given by different dealing rules and sampling techniques in the four plots are similar to each other. In other words, the simulation results are independent of the numbers of the agents, total amount of money, and temporal steps for Rule I, Rule II, Sampling I and Sampling II, respectively. It is demonstrated that the kinetic models with the transaction rules and sampling techniques own high numerical robustness and reliability. 

\section{Numerical investigation}\label{SecIV}
\begin{figure}[htbp]
	\centerline{\includegraphics[width=5 cm]{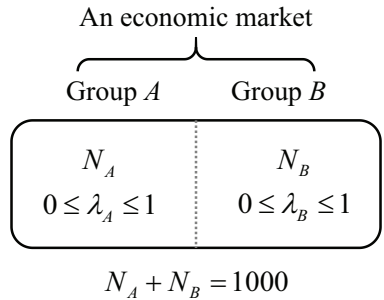}}
	\vspace*{8pt}
	\caption{Sketch of two groups of agents in an economic market.\label{Fig08}}
\end{figure}

In this section, let us study the features of the stationary wealth distributions of agents with or without saving propensities in the economic society. To this end, we assume that there are two groups ${A}$ and ${B}$ in a financial market, see Fig. \ref{Fig08}. The numbers of agents in the two groups are ${N}_{A}$ and ${N}_{B}$, and the ranges of the corresponding saving propensities are $0 \le {\lambda}_{A} \le 1$ and $0 \le {\lambda}_{B} \le 1$, respectively. After plenty of transactions, the commercial system reaches the steady state. 

\subsection{Impact of saving propensities}

First of all, the influence of saving propensities upon the wealth distribution is investigated. For convenience, the numbers of agents in groups ${A}$ and ${B}$ are set as ${N}_{A} = {N}_{B} = 500$, and the saving propensities of agents in the two groups increase from zero to one. To quantitatively study the relationship between saving propensities and wealth inequality, three useful parameters (i.e., the Gini coefficient $\rm{G}$, Kolkata index $\rm{k}$, and deviation degree $\Delta$) are employed in this part. 

\begin{figure}[htbp]
	\centerline{\includegraphics[width=12.7cm]{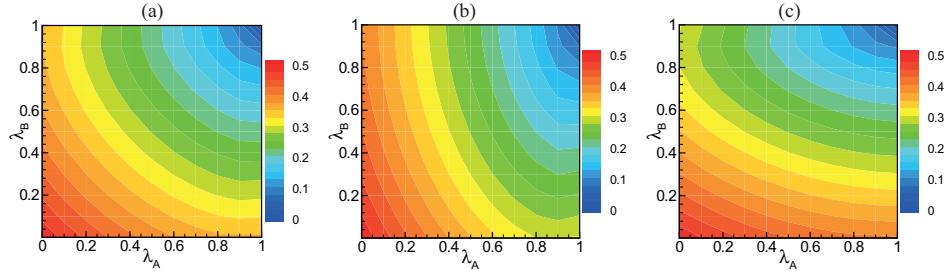}}
	\vspace*{8pt}
	\caption{Contours of Gini coefficients in the space of saving propensities ${\lambda}_{A}$ and ${\lambda}_{B}$: (a) the whole system, (b) group $A$, and (c) group $B$. \label{Fig09}}
\end{figure}

Figure \ref{Fig09} depicts the Gini coefficients versus the saving propensities ${\lambda}_{A}$ and ${\lambda}_{B}$. Figures \ref{Fig09} (a)-(c) show the Gini coefficients for individual agents in the whole economic market, group $A$, and group $B$, respectively. The values of Gini coefficients are labelled in the legend. On the whole, the contours of Gini coefficients decrease with the increasing of ${\lambda}_{A}$ and/or ${\lambda}_{B}$. In each plot, the minima and maxima of saving propensities are $0$ and $0.5$, respectively. That is to say, the wealth distribution tends to be even as the saving propensities increases. 
The increasing savings can reduce the inequality of wealth income distribution and inhibit the polarization of wealth. To be specific, without any saving (i.e., ${\lambda_i} = 0$), a relatively larger amount of money could be exchanged among traders, and the resulting wealth difference between agents could be greater. On the contrary, for the maximum saving propensity (i.e., ${\lambda_i} = 1$), no transaction takes place, hence each agent keeps identical wealth and the Gini index is zero, as earlier. 
Additionally, the contours are symmetrical about the diagonal line in Fig. \ref{Fig09} (a), and are asymmetric about the diagonal line in Fig. \ref{Fig09} (b) or (c). 

\begin{figure}[htbp]
	\centerline{\includegraphics[width=12.7cm]{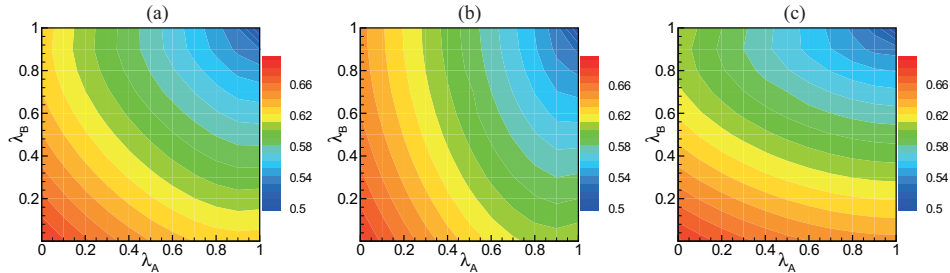}}
	\vspace*{8pt}
	\caption{Contours of Kolkata indices in the space of saving propensities ${\lambda}_{A}$ and ${\lambda}_{B}$: (a) the whole system, (b) group $A$, and (c) group $B$. \label{Fig10}}
\end{figure}

Besides, the Kolkata index is another important parameter to measure the wealth inequality. Next, let us study the Kolkata indices when the economic system reaches the steady state. Figure \ref{Fig10} illustrates the Kolkata indices versus the saving propensities ${\lambda}_{A}$ and ${\lambda}_{B}$. It is obvious in Figs. \ref{Fig10} (a)-(c) that the Kolkata indices vary from $0.5$ to $0.68$, and become smaller for greater saving propensities ${\lambda}_{A}$ and/or ${\lambda}_{B}$. Similar to the Gini coefficient, the contours of Kolkata indices are also symmetrical about the diagonal line in Fig. \ref{Fig10} (a), and are asymmetric about the diagonal line in Fig. \ref{Fig10} (b) or (c). 
Additionally, it can be found in Figs. \ref{Fig09} and \ref{Fig10} that, with the increasing saving propensities, both Gini coefficients and Kolkata indices reduce, namely, the wealth inequality is weakened in the social system. 

\begin{figure}[htbp]
	\centerline{\includegraphics[width=12.7cm]{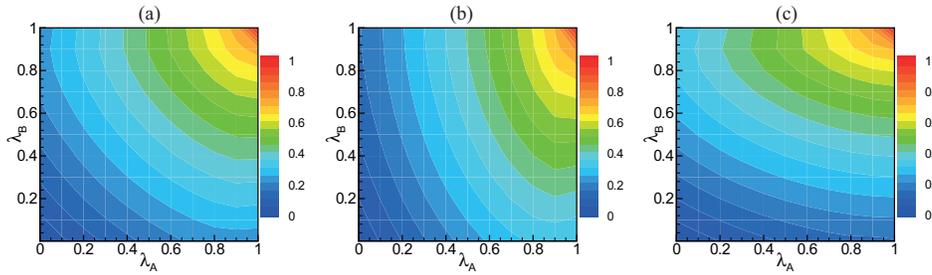}}
	\vspace*{8pt}
	\caption{Contours of deviation degrees in the space of saving propensities ${\lambda}_{A}$ and ${\lambda}_{B}$: (a)  the whole system, (b) group $A$, and (c) group $B$. \label{Fig11}}
\end{figure}

Next, to describe the difference of the wealth distributions of individual agents with and without saving propensities, let us introduce the definition of deviation degree \cite{Cui2021PA},
\begin{equation}
	\Delta
	=\frac{\int_{0}^{\infty }{\left| f-{{f}_{0}} \right|}dm}{\int_{0}^{\infty }{\left| f+{{f}_{0}} \right|}dm}
	\tt{,}
\label{Deltaf_Defination}
\end{equation}
where $f = P_1 (\lambda_A, \lambda_B)$ denotes the wealth distribution of individual agents with or without saving propensities, and ${f}_{0} = P_1 (0, 0)$ stands for the wealth distribution without saving propensities. 
The integral of probability $f$ or ${f}_{0}$ is equal to one, namely, 
\begin{equation}
	\int_{0}^{\infty }{f } dm = 1
	\tt{,}
\end{equation}
\begin{equation}
	\int_{0}^{\infty }{{f}_{0} } dm = 1
	\tt{,}
\end{equation}
therefore Eq. \ref{Deltaf_Defination} can be written as 
\begin{equation}
	\Delta = \frac{1}{2} \int_{0}^{\infty }{\left| f - {f}_{0} \right|} dm
	\tt{,}
	\label{Deltaf}
\end{equation}
In theory, the realm is $0 \le \Delta \le 1 $, and the departure of the distribution $f$ from ${f}_{0}$ increases as the deviation degree $\Delta$ increases from zero to one. 

Figure \ref{Fig11} plots the deviation degrees versus the saving propensities ${\lambda}_{A}$ and ${\lambda}_{B}$. Figures \ref{Fig11} (a)-(c) illustrate the deviation degrees of the whole economic system, group $A$, and group $B$, respectively. Clearly, the deviation degree varies from zero to one in each plot. In contrast to the Gini coefficients and Kolkata indices, the deviation degree becomes large for high saving propensities ${\lambda}_{A}$ and/or ${\lambda}_{B}$. Similar to Gini coefficients and Kolkata indices, the contours of deviation degrees are also symmetrical about the diagonal line in Fig. \ref{Fig11} (a), and are asymmetric about the diagonal line in Fig. \ref{Fig11} (b) or (c). 

It is noteworthy that the deviation degree can be employed to measure the cooperatively self-organizing manifestation in the market under the influence of saving propensities. 
Actually, Eqs. \ref{Probability1}-\ref{Lorenz2} no longer hold for any nonvanishing saving (${{\lambda }_{i}}\ne 0$) in the market. The equilibrium distribution becomes the asymmetric Gaussian-like with the most probable money per agent shifting away from zero to the mean wealth as the saving propensity increases from $\lambda =0$ to $\lambda \to 1$. For larger savings, there are fewer paupers and rich persons, and more agents possess wealth close to the average. This global feature, resulting from the individual self-interest of saving a portion of wealth, could be regarded as a demonstration of the self-organization in the market \cite{Chakraborti2000EPJB,Chatterjee2004PA}. In fact, the cooperatively self-organizing characteristics, induced by sheer self-interest of saving by each agent, are quite significant in the social system \cite{Chakraborti2000EPJB,Chatterjee2004PA}. 
In sum, with increasing of saving propensities, the wealth distribution function in the cooperatively interacting market (${{\lambda }_{i}}\ne 0$) departs far from the Boltzmann-Gibbs form in the ideal free market (${{\lambda }_{i}}=0$), thus the corresponding deviation degree becomes large. 

In addition, it should be pointed out that the Gini coefficients, Kolkata indices, deviation degrees are functions of the saving propensities ${\lambda}_{A}$ and ${\lambda}_{B}$, as shown in Figs. \ref{Fig09} - \ref{Fig11}. The three parameters describe the wealth inequality from different perspectives. It can be found that the wealth inequality is weakened as the saving propensities ${\lambda}_{A}$ and/or ${\lambda}_{B}$ rise from zero to one. In addition, the contours for groups $A$ and $B$ are the same if the axes ${\lambda}_{A}$ and ${\lambda}_{B}$ are transposed in Figs. \ref{Fig09} - \ref{Fig11}, because groups $A$ and $B$ own the same number of agents.

\subsection{Impact of proportions of agents}

In fact, the proportions of agents with different saving propensities make a significant impact on the wealth distribution in the realistic society. Next, we probe the effects of proportions of agents with different saving propensities in groups $A$ and $B$. Without loss of generality, the total number of agents is fixed as $N = {N}_{A} + {N}_{B} = 1000$. The proportions of the agents in the two groups are adjustable. Let us introduce the symbol, ${F_A} = {{N}_{A}} / N$, which stands for the proportion of agents in group $A$. Hence the proportion of agents in group $B$ is ${F_B} = 1 - {F_A}$. Then, the values of ${F_A}$ and ${F_B}$ can be adjusted from zero to one and from one to zero, respectively. Moreover, for brevity, only three typical cases with different saving propensities (${\lambda}_{A}$, ${\lambda}_{B}$) $=$ ($0.0$, $1.0$), ($0.2$, $0.8$), and ($0.4$, $0.6$), are under consideration in this subsection. In other words, the sum of the saving propensities of the two groups is fixed as ${\lambda}_{A} + {\lambda}_{B} = 1$. More interesting financial circumstances are beyond this paper. 

\begin{figure}[htbp]
	\centerline{\includegraphics[width=12.7cm]{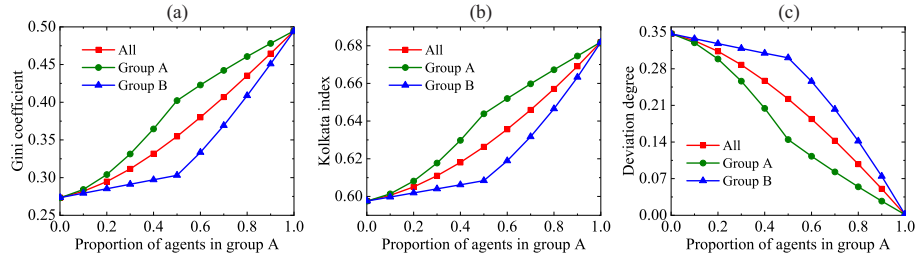}}
	\vspace*{8pt}
	\caption{The Gini coefficients (a), Kolkata indices (b), and deviation degrees (c) versus the proportion of agents in group $A$. The saving propensities are ${\lambda}_{A}=0.0$ and ${\lambda}_{B}=1.0$. The lines with squares, circles and triangles represent the whole system, group $A$ and group $B$, respectively.\label{Fig12}}
\end{figure}

Figure \ref{Fig12} delineates the Gini coefficients (a), Kolkata indices (b), and deviation degrees (c) versus the proportion of agents in group $A$ when the commercial market reaches the steady state. From Fig. \ref{Fig12}, the following points can be obtained. 

(i) On the whole, with the increasing number of agents in group $A$, both the Gini coefficients and Kolkata indices increase monotonously, while the deviation degrees decrease monotonously. The Gini coefficients of group $A$ ($B$) are higher (lower) than that of all agents under the condition $0 < {F}_{A} < 1$. Besides, the Kolkata indices own similar features, while the deviation degrees show opposite tendencies. 

(ii) As shown in Fig. \ref{Fig12} (a), the Gini coefficients of the whole system, group $A$, and group $B$ become close to the minimum $\rm{G} = 0.27$ as the proportion ${F}_{A}$ approaches zero, and are near the maximum $\rm{G} = 0.5$ as ${F}_{A}$ tends towards one. Similarly, the range of the Kolkata indices is $0.597 \le \rm{k} \le 0.682$ in Fig. \ref{Fig12} (b). On the contrary, the deviation degrees tend to the maximum ${\Delta} = 0.35$ as ${F}_{A} \to 0$, and the minimum ${\Delta} = 0$ as ${F}_{A} \to 1$. 

(iii) The patterns of Gini coefficients, Kolkata indices, and deviation degrees of all agents show smooth curves, while these parameters in group $A$ or $B$ have an inflection point located at the abscissa ${{F}_{A}} = 0.5$, i.e., ${{N}_{A}} = {{N}_{B}}$. Specifically, the slope of the Gini coefficient in group $A$ for ${{N}_{A}} < {{N}_{B}}$ is larger than that for 
	${{N}_{A}} > {{N}_{B}}$. 
On the contrary, the slope of the Gini coefficient in group $B$ increases significantly at the connection point
 ${{F}_{A}} = 0.5$. Furthermore, the Kolkata indices (deviation degrees) have similar (opposite) variations. 

\begin{figure}[htbp]
	\centerline{\includegraphics[width=12.7cm]{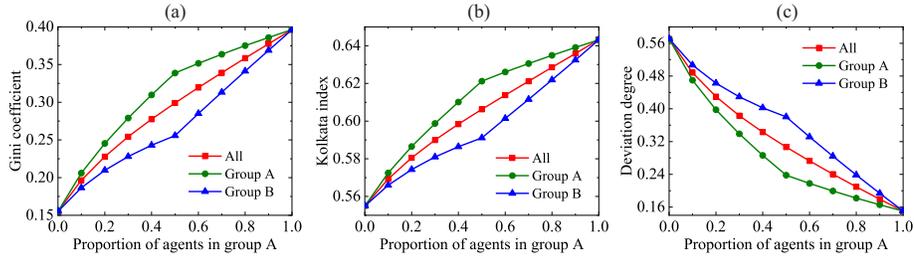}}
	\vspace*{8pt}
	\caption{The Gini coefficients (a), Kolkata indices (b), and deviation degrees (c) versus the proportion of agents in group $A$. The saving propensities are ${\lambda}_{A}=0.2$ and ${\lambda}_{B}=0.8$. The lines with squares, circles and triangles represent the whole system, group $A$ and group $B$, respectively.\label{Fig13}}
\end{figure}

\begin{figure}[htbp]
	\centerline{\includegraphics[width=12.7cm]{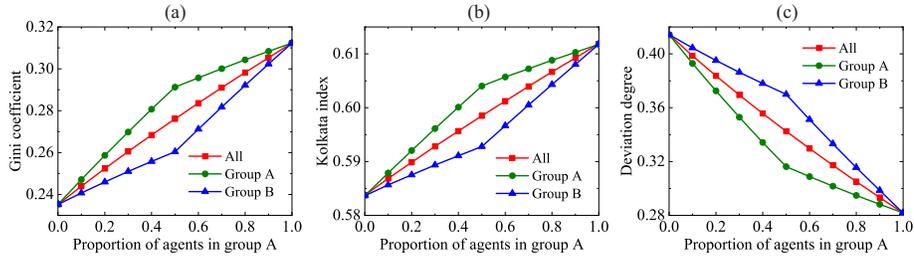}}
	\vspace*{8pt}
	\caption{The Gini coefficients (a), Kolkata indices (b), and deviation degrees (c) versus the proportion of agents in group $A$. The saving propensities are ${\lambda}_{A}=0.4$ and ${\lambda}_{B}=0.6$. The lines with squares, circles and triangles represent the whole system, group $A$ and group $B$, respectively.\label{Fig14}}
\end{figure}

To have a deeper understanding of the impact of proportions of agents on the wealth distribution, we make a comparison of simulations with various saving propensities. 
Figures \ref{Fig13} and \ref{Fig14} illustrate the simulation results with saving propensities $({\lambda}_{A}, {\lambda}_{B}) = (0.2, 0.8)$ and $(0.4, 0.6)$, respectively. It can be found that Figs. \ref{Fig12} - \ref{Fig14} have the following three similarities. 

(i) Both Gini coefficients and Kolkata indices increase monotonously, and the deviation degrees decrease monotonously, as the proportion of agents in group $A$ becomes large. Meanwhile, in the case of $0 < {F}_{A} < 1$, both Gini coefficients and Kolkata indices of group $A$ ($B$) are higher (lower) than that of the whole system, while the deviation degrees show opposite tendencies. 

(ii) All lines of Gini coefficients, Kolkata indices and deviation degrees of the whole system, group $A$ and group $B$ converge at the starting point (${F}_{A} = 0$) and end point (${F}_{A} = 1$). To be specific, the starting (end) points correspond to the minima (maxima) of Gini coefficients and Kolkata indices, and the maxima (minima) of deviation degrees. 

(iii) The lines of Gini coefficients, Kolkata indices or deviation degrees of all agents are smooth in the whole range, while each curve for group $A$ or $B$ has a inflection point located at the abscissa ${{F}_{A}} = 0.5$, and the changing trends across the inflection point are similar for heterogeneous saving propensities. For example, the left limit of the slope of the Gini coefficient of group $A$ is higher than the right limit at the inflection point. The Kolkata indices (deviation degrees) have identical (opposite) characteristics. 

In addition to the aforementioned similarities, there are some differences among Figs. \ref{Fig12} - \ref{Fig14}. 

(i) The ranges (as well as the minima and maxima) of the Gini coefficients, Kolkata indices or deviation degrees are different under various conditions of saving propensities. For example, the ranges of the Gini coefficients are $0.27 \le \rm{G} \le 0.5$, $0.15 \le \rm{G} \le 0.40$, and $0.23 \le \rm{G} \le 0.32$ in Figs. \ref{Fig12} - \ref{Fig14}, respectively. So do the Kolkata indices and deviation degrees. 

(ii) The variations of Gini coefficients, Kolkata indices or deviation degrees are different for diverse saving propensities ${\lambda}_{A}$ and ${\lambda}_{B}$. Moreover, the slope changes of Gini coefficients, Kolkata indices or deviation degrees around the connection point are different. For instance, the left (right) limits of the slopes of the Gini coefficients of group $A$ are different with each other in Figs. \ref{Fig12} - \ref{Fig14}. 

\section{Conclusion}\label{SecV}
In this work, we propose the standard and reduced gas automata for a closed economic market, where the total amount of money and the number of agents are fixed, and the saving propensities of agents can be either homogeneous or inhomogeneous. The reduced LGA, which is a  more simplified kinetic model, is a special version of the standard LGA excluding the propagation phase. Consequently, this simplified econophysics model possesses a higher computational efficiency than the standard LGA. Besides, two types of transaction rules are under consideration. For rule I, the trading results are a function of the mean saving propensities of two agents who get engaged in a trade. For rule II, the business is controlled by a random parameter between the saving propensities of two traders. Additionally, two sampling methods are introduced for the random selection of two agents in trade. Specifically, Sampling I is the sampling with replacement and is easier to program due to its simplicity. Sampling II is the sampling without replacement and takes less time to produce the equilibrium wealth distribution because of its high computing efficiency. 

To validate the standard LGA and the simplified kinetic model with aforementioned dealing rules and sampling approaches, we measure both nonequilibrium and equilibrium states of the financial systems where two groups of agents have the same or different saving propensities. On the one hand, the relaxation processes are similar for Rules I and II, and different for Samplings I and II. On the other hand, there are slight differences between the steady wealth distributions obtained from the two transaction rules and two sampling techniques. For both the standard and reduced lattice gas automata, the law of money conservation is always obeyed in the evolution of the economic market. Furthermore, under the condition without saving propensities, the simulation results coincide with the theoretical solutions of the wealth probabilities or Lorenz curves \cite{Dragulescu2001}. Besides, in the case with or without homogeneous saving propensities, our rules and samplings give identical simulation data to those of the CC method \cite{Chakraborti2000EPJB}. In addition, the numerical robustness and reliability are verified by using different numbers of agents, values of average money, and time instants.

Moreover, we investigate the influence of saving propensities upon the wealth distribution. The contours of Gini coefficients, Kolkata indices and deviation degrees of the whole market and two groups are plotted in the space of saving propensities ${\lambda}_{A}$ and ${\lambda}_{B}$. On the whole, the Gini coefficients and Kolkata indices decrease with the increasing of ${\lambda}_{A}$ and/or ${\lambda}_{B}$, while the deviation degrees become large for high saving propensities ${\lambda}_{A}$ and/or ${\lambda}_{B}$. That is to say, the wealth inequality is weakened as the saving propensities become large. Finally, we probe the effects of proportions of agents with different saving propensities in two groups. Three cases of saving propensities are simulated under the condition ${\lambda}_{A} < {\lambda}_{B}$. It is found that both Gini coefficients and Kolkata indices increase monotonously, and the deviation degrees decrease monotonously, as the proportion of agents in group $A$ becomes large. The ranges and variations of the Gini coefficients, Kolkata indices or deviation degrees are different for various saving propensities.

\appendix

\section{} \label{AppendixA}

\begin{figure}[htbp]
	\centerline{\includegraphics[width=12.7cm]{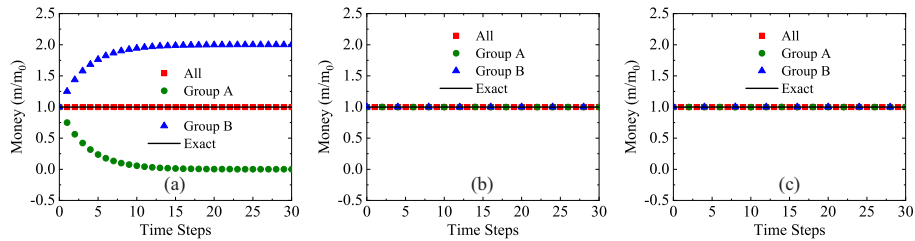}}
	\vspace*{8pt}
	\caption{Evolution of average money: (a) CCM, (b) Rule I, (c) Rule II. The squares, circles, and triangles represent average money of the whole system, group $A$, and group $B$, respectively. The solid lines stand for the exact solutions. \label{Fig15}}
\end{figure}

In this paper, we propose two types of transaction rules that are different from the CCM model in Ref.~\refcite{Chatterjee2004PA}. 
To make a comparison between them, Figs. \ref{Fig15} (a)-(c) illustrates the evolution of average money given by the CCM, Rule I, and Rule II, respectively. The saving propensities are ${\lambda}_{A} = 0.0$ and ${\lambda}_{B} = 1.0$, and other parameters are the same as those in Fig. \ref{Fig04}. The squares, circles, and triangles indicate the mean money of all agents, group A and group B, respectively. The solid lines stand for the exact solutions $m = {m_0}$. In theory, there is (no) exchange of money between agents with saving propensities ${\lambda}_{A} = 0.0$ (${\lambda}_{B} = 1.0$) in group $A$ ($B$). 

Figure \ref{Fig15} (a) shows that the average money of group $A$ decreases gradually and finally becomes zero, while the average money of group $B$ increases and ultimately reaches two. That is to say, the agents of group $A$ will have no money, and the agents of group $B$ will possess all the money in the equilibrium state. Obviously, this is not a logical conclusion. In contrast, Figs. \ref{Fig15} (b) and (c) show that, by using Rules I or II, the average money of agents in the whole system, group $A$ or $B$ is close to the exact solutions. Consequently, it is concluded that our model is more sensible and realistic than the CCM model in Ref.~\refcite{Chatterjee2004PA}. 

\section*{Acknowledgments}

This work is supported by the National Natural Science Foundation of China (under Grant No. 51806116), and Guangdong Basic and Applied Basic Research Foundation (under Grant No. 2022A1515012116).

\bibliographystyle{ws-ijmpc}

\bibliography{References}

\end{document}